\documentclass[twocolumn,english,preprintnumbers,amsmath,amssymb,prb]{revtex4}

\usepackage{graphicx}
\usepackage{color}

\begin{document}

\title{Quantum effects in structural and elastic properties of graphite:
       Path-integral simulations}
\author{Carlos P. Herrero}
\author{Rafael Ram\'irez}
\affiliation{Instituto de Ciencia de Materiales de Madrid,
         Consejo Superior de Investigaciones Cient\'ificas (CSIC),
         Campus de Cantoblanco, 28049 Madrid, Spain }
\date{\today}

\begin{abstract}
Graphite, as a well-known carbon-based solid, is a paradigmatic 
example of the so-called van der Waals layered materials, which
display a large anisotropy in their physical properties.
Here we study quantum effects in structural and elastic
properties of graphite by using path-integral molecular 
dynamics simulations in the temperature range from
50 to 1500~K.  This method takes into account quantization 
and anharmonicity of vibrational modes in the material.
Our results are compared with those found
by using classical molecular dynamics simulations.
We analyze the volume and in-plane area
as functions of temperature and external stress.
The quantum motion is essential to correctly describe
the in-plane and out-of-plane thermal expansion. 
Quantum effects cause also changes in the elastic
properties of graphite with respect to a classical model.
At low temperature we find an appreciable decrease in the
linear elastic constants, mainly in $C_{12}$ and $C_{44}$.
Quantum corrections in stiffness constants can be in some 
cases even larger than 20\%.
The bulk modulus and Poisson's ratio are reduced in a 4\% and 
19\%, respectively, due to zero-point motion of the C atoms.
These quantum effects in structural and elastic properties 
of graphite are nonnegligible up to temperatures higher 
than 300~K.
\end{abstract}

\maketitle

\section{Introduction}

Over the last few decades, we have witnessed a large progress in 
the knowledge of carbon-based materials with $sp^2$ orbital 
hybridization, as fullerenes, carbon nanotubes, and 
graphene,\cite{ho01,ho00,ge07,ka07} which has progressively 
broadened the scope of this research field further than 
the traditionally known graphite.
This classical material has in turn become a paradigmatic 
case of the nowadays called van der Waals materials, 
characterized by a layered structure, where the interactions 
between sheets are much weaker than those between atoms in each
sheet.\cite{fr20}  Among these materials one finds hexagonal 
boron nitride, transition-metal dichalcogenides, and III-VI 
compounds as InSe and GaS.\cite{aj16}

 In addition to its common role in several areas as lubrication, 
batteries, and nuclear technology, graphite has had a renewed 
interest in recent years in connection with the discovery 
of graphene and the potential applications of this 
two-dimensional material.   In particular, mechanical properties 
of graphite, including elastic constants, have been studied 
by using experimental and theoretical 
methods.\cite{bl70,bo07b,ni72,bo97,ja87,ha04,mo05,mi08b,sa11}
Nevertheless, a precise knowledge of these properties
has been limited by the difficulty of obtaining 
high-quality single crystalline samples.\cite{ro06,zh05}
Thus, although there is a general agreement on the values of the
largest elastic stiffness constants (i.e., their relative uncertainty
is a few percent), values of smaller elastic constants such as 
$C_{44}$ and $C_{13}$ are known with relatively large error bars. 
For $C_{13}$, for example, one finds in the literature values of 
0(3) GPa\cite{bo07b} and 15(5) GPa,\cite{bl70}
derived from apparently reliable methods.
This is in part due to the high anisotropy of graphite, which
means that elastic stiffness constants related to in-plane deformations,
as $C_{11}$ and $C_{12}$, are much larger than those related
to deformations along the $z$-axis perpendicular to the 
basal plane.

Theoretical work has been carried out to study structural,
elastic, and thermodynamic properties of graphite.
Most of the calculations and simulations performed to analyze
such properties of graphite (and solids in general)
have considered atomic nuclei as classical particles. 
This means that their quantum zero-point motion is not 
taken into account in zero-temperature calculations, and
their motion is assumed to be classical (i.e., follows Newton's laws)
in finite-temperature Monte Carlo or molecular dynamics (MD) simulations.
The quantum delocalization of atomic nuclei becomes unimportant
at high temperatures, but can lead to appreciable corrections 
in physical observables for $T$ lower than the Debye temperature 
of the material, $\Theta_D$.\cite{ki96}
Throughout this paper we call nuclear quantum effects those caused
by the quantum nature of atomic nuclei, which manifests itself
in a spacial delocalization larger than that expected for 
a classical calculation (thermal motion).

Several research groups employed density-functional theory (DFT)
calculations at $T = 0$, and in some cases finite temperatures
were considered by using a quantum quasiharmonic approximation (QHA) 
for the vibrational modes.\cite{mo05,se14,ma17}
This approach is generally accepted to be sound at low temperature, but
it can be inaccurate for layered materials at relatively high temperatures,
as a consequence of
an appreciable anharmonic coupling between out-of-plane and
in-plane vibrational modes, which is not considered in a QHA.
Various works based on classical molecular dynamics 
and Monte Carlo simulations of graphite have also appeared in the 
literature.\cite{gh05b,ts10,co11,tr16,pe18,ko14b,du11}
For this layered material, frequencies of out-of-plane vibrational
modes are lower than those of in-plane vibrations, and one can define
two different Debye temperatures, one for the first set of modes 
($\Theta_D^{\rm out} \sim$~1000~K) and another for 
the second ($\Theta_D^{\rm in} \sim$ 2500~K).\cite{kr53,ni03} 
This means that nuclear quantum effects are expected to be
appreciable at temperatures in the order of 300~K and even higher.

The difficulties associated to using classical simulations
can be surmounted by employing simulation techniques
which take account of nuclear quantum effects in an explicit way, as
those based on Feynman path integrals.\cite{gi88,ce95,br15,he16}
This procedure is in principle equivalent to a quantization of
the vibrational modes in the solid, with the advantage that
anharmonicities are directly included in the path-integral 
simulation procedures. This kind of methods have been used to
study properties of materials as diamond,\cite{he01,br20} 
silicon,\cite{no96} boron nitride,\cite{ca16,br19} 
and graphene.\cite{br15,he16,he19}
We are not aware of any quantum atomistic simulation of graphite.
An important point is the large anisotropy of this material,
so that quantum effects can be quantitatively very different
for properties along directions on the basal plane or 
perpendicular to it.

Here we employ the path-integral molecular dynamics (PIMD) 
method to study structural and elastic properties of graphite 
in a temperature range from 50 to 1500~K.
The importance of nuclear quantum effects in the considered variables is
assessed by comparing the results of quantum simulations with
those obtained from classical MD simulations.
We find that considering nuclear quantum motion is necessary for
an adequate description of the in-plane thermal expansion.
In general, quantum effects are nonnegligible in structural and
elastic properties of graphite for temperatures even higher than 300~K.
Particular attention is set on the temperature dependence of 
the linear elastic constants and bulk modulus of graphite.
At low temperature, quantum corrections in elastic stiffness 
constants may be higher than 20\%, whereas the
Poisson's ratio and bulk modulus are appreciably reduced.

The paper is organized as follows. In Sec.~II we describe the
computational methods employed in the simulations.
In Sec.~III we discuss the phonon dispersion bands and the 
calculation of elastic constants at $T = 0$.
Results for the internal energy of graphite are presented
in Sec.~IV. In Sec.~V we show results for the volume and the 
in-plane area, and the thermal expansion is presented in Sec.~VI.
Data of the elastic constants and bulk modulus at finite temperatures 
are given and discussed in Secs.~VII and VIII.
Finally, we summarize the main results in Sec.~IX.

\section{Computational Method}

In this paper we study the influence of nuclear quantum effects on
structural and elastic properties of graphite. This means that
we consider quantum delocalization of atomic nuclei, and analyze its
influence on physical observables of the material.
This requires, on one side, the definition of a reliable
potential to describe the interatomic interactions in the solid.
This potential is usually derived from {\em ab-initio} methods
(e.g., DFT), tight-binding-like Hamiltonians, or effective 
interactions. This provides one with a Born-Oppenheimer surface
for motion of the atomic nuclei.
On the other side, we need a method to take into account the quantum
dynamics (or quantum delocalization) in the many-body configuration
space of atomic coordinates with the selected interatomic
interactions. This means that we have to base our finite-temperature 
calculations on quantum statistical physics, in contrast to 
the more usually employed classical statistical physics to
perform Monte Carlo or molecular dynamics simulations.

Thus, we employ PIMD simulations to study equilibrium properties of
graphite as a function of temperature and pressure.
The PIMD method rests on the Feynman path-integral formulation of
statistical mechanics,\cite{fe72} which turns out to be 
a suitable nonperturbative procedure
to study many-body quantum systems at finite temperatures.
In the implementation of this method,
each quantum particle (here, atomic nucleus) is described as 
a set of $N_{\rm Tr}$
(Trotter number) beads, which act as classical particles 
building a ring polymer.\cite{gi88,ce95}
In this way, one has a {\em classical isomorph} displaying an
unreal dynamics, as it does not represent the true dynamics of
the actual quantum particles. This isomorph is, however, 
practical for an efficient sampling of the configuration space, 
thus giving accurate values for
time-independent variables of the quantum system.
Details on this simulation technique can be found
in Refs.~\onlinecite{gi88,ce95,he14,ca17}.

Interatomic interactions between C atoms are described here
through a long-range bond order potential, the so-called LCBOPII,
mainly employed earlier to perform classical simulations 
of carbon-based systems.\cite{lo05}
Notably, it has been used to study the phase diagram of carbon,
including graphite, diamond, and the liquid, and displayed its 
precision by yielding rather accurately the graphite-diamond
transition line.\cite{gh05b}
More recently, this effective potential has been also found
to accurately describe various properties 
of graphene.\cite{fa07,lo16,za09,po12,ra17}

The LCBOPII potential was also employed in last years to perform 
PIMD simulations, providing a quantification of nuclear quantum effects
in monolayer and bilayer graphene from a comparison with results 
of classical simulations.\cite{he16,he18}
In this paper about graphite, as in earlier simulations of 
graphene,\cite{ra16,he16,ra17} the original parameterization 
of the LCBOPII potential has been slightly modified to increase 
the zero-temperature bending constant $\kappa$ of the graphene
layers from 1.1 eV to 1.49 eV, closer to experimental 
data.\cite{la14,ti17}
The interlayer interaction was fitted to the results of quantum 
Monte Carlo calculations,\cite{sp09} to yield a binding energy of 
50 meV/atom for graphite.\cite{za10b}

For the calculations presented here, we have employed both
the isothermal-isobaric ($NPT$) and isothermal-isochoric ($NVT$)
ensembles.  For the $NVT$ simulations we used cell parameters
obtained from equilibrium $NPT$ simulations at the same
temperature.
We have used effective algorithms for the PIMD simulations,
as those presented in the literature.\cite{tu92,tu98,ma99}
In particular, we have employed staging coordinates to define 
the bead positions in the classical isomorph, and in order 
to keep a constant $T$ we have introduced chains of four 
Nos\'e-Hoover thermostats connected to each staging coordinate.
In $NPT$ simulations, another chain of four thermostats was 
coupled to the barostat to give the equilibrium volume 
fluctuations for the considered external stress.\cite{tu10,he14}.
The equations of motion have been integrated by using the reversible
reference system propagator algorithm (RESPA), which allows to consider
different time steps for the integration of the slow and fast
degrees of freedom.\cite{ma96}
The time step corresponding to the interatomic forces was
$\Delta t$ = 1 fs, which is adequate for 
the C atomic mass and the range of temperatures considered here.
More details on this type of PIMD simulations are given
elsewhere.\cite{tu98,he06,he11}

We considered orthorhombic simulation cells of graphite with 
$N$ = 960 atoms and similar side lengths in the in-plane $x$ 
and $y$ directions ($L_x \approx L_y$).
These cells included four carbon sheets in AB stacking, each 
with $n$ = 240 atoms.  Periodic boundary conditions were assumed.
To check the convergence of our results, some simulations were
carried out for larger simulation cells with $n$ = 960 atoms.
As the size $n$ is increased, there appear vibrational modes
with longer wavelength $\lambda$. In fact, one has an effective
wavelength cutoff 
$\lambda_{\rm max} \approx L$, where $L = (L_x L_y)^{1/2}$,
which translates into a wavevector cutoff
$k_{\rm min} \approx 2 \pi / L$, with $k = |{\bf k}|$.
The results obtained using $n$ = 240 and 960 atoms/layer 
for the energy, in-plane area, and interatomic distances coincide 
within the statistical error bars of our simulations.
For example, for the energy and mean interatomic distance, differences
are less than $6 \times 10^{-4}$ eV/atom and $4 \times 10^{-5}$ \AA,
respectively.

Sampling of the configuration space was performed in the temperature
range between 50~K and 1500~K.
The Trotter number $N_{\rm Tr}$ (number of beads in the ring polymers)
varies with the temperature as $N_{\rm Tr} = 6000~{\rm K} / T$,
which gives a roughly constant accuracy of the PIMD results for
different temperatures.\cite{he06,he11,ra12}
A typical simulation run in the $NVT$ or $NPT$ ensembles consisted
of $2 \times 10^5$ PIMD steps for system equilibration and
$8 \times 10^6$ steps for calculation of average variables.
For comparison with the results of our quantum PIMD simulations,
we have also carried out classical molecular dynamics simulations
with the same interatomic potential. In our context, 
these classical simulations correspond to a Trotter number
$N_{\rm Tr} = 1$. 

\begin{figure}
\vspace{-0.6cm}
\includegraphics[width=7cm]{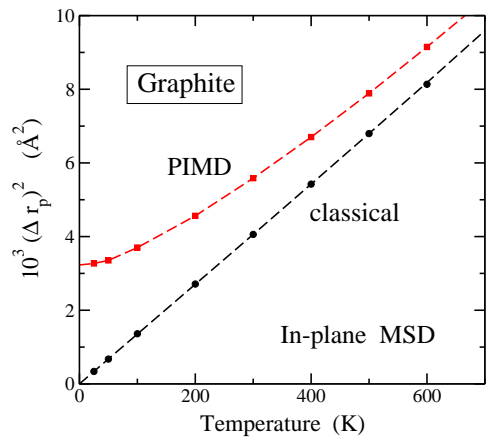}
\vspace{-0.5cm}
\caption{In-plane mean-square displacement,
$(\Delta r_p)^2 = (\Delta x)^2 + (\Delta y)^2$, of carbon atoms
in graphite, as derived from classical MD (circles) and quantum
PIMD simulations (squares) at various temperatures.
Error bars are less than the symbol size.
}
\label{f1}
\end{figure}

In Fig.~1 we present the mean-square displacement (MSD) of carbon
atoms in the $(x, y)$ layer plane, 
$(\Delta r_p)^2 = (\Delta x)^2 + (\Delta y)^2$,
at several temperatures and zero external pressure.
Solid circles represent data points obtained from classical MD
simulations, whereas squares indicate results of PIMD simulations.
The classical results converge to zero in the low-temperature
limit, as expected in classical physics, while the quantum
data converge at low $T$ to 
$(\Delta r_p)^2 = 3.4 \times 10^{-3}$ \AA$^2$
(in-plane zero-point delocalization). The difference between 
classical and quantum results decreases as temperature is raised, 
but is clearly appreciable in the whole temperature range
displayed in Fig.~1. An even larger quantum delocalization occurs for
the out-of-plane $z$-direction, which has been studied in detail for 
graphene in Ref.~\onlinecite{he16}. Such atomic quantum delocalization
(in-plane and out-of-plane) causes changes in the properties of
the material, especially in the presence of anharmonicities in
the lattice vibrations, as the atomic motion "explores" larger
regions of the configuration space, as compared to classical
simulations.

The elastic stiffness constants at $T = 0$ have been calculated
from the phonon dispersion bands and cell distortions as explained
in Sec.~III.
Relations between the elastic stiffness constants $C_{ij}$ and 
compliance constants $S_{ij}$ for hexagonal crystals,
as well as their definitions as functions of the strain and stress
components, $e_{ij}$ and $\tau_{ij}$, are given in 
the literature\cite{ma18,li90,ra20b}.
We use the standard notation for strain components,
with $e_{ij} = \epsilon_{ij}$ for $i = j$, and
$e_{ij} = 2 \epsilon_{ij}$ for $i \neq j$ .\cite{as76,ma18}

At finite temperatures, we have also obtained the elastic constants 
of graphite in two different ways.
The first way consists in applying a certain component of the 
stress tensor in isothermal-isobaric simulations, and obtaining
the associated elastic constants from the resulting strain.
Thus, for example, for $\tau_{xx} \neq 0$ and $\tau_{ij} = 0$
for the other components we can calculate $S_{11}$, $S_{12}$,
and $S_{13}$. Then, from the obtained compliance constants 
we calculate the stiffness constants $C_{ij}$ using the relations 
corresponding to hexagonal crystals.\cite{ma18,li90,ra20b}

In the second way, we take as a reference for each temperature 
and kind of simulation (classical MD or PIMD) the simulation
cell parameters obtained from equilibrium isothermal-isobaric
simulations at that temperature. Then, we carry out $NVT$
simulations for cells strained a certain amount respect the
equilibrium one. For example, for $e_{zz} \neq 0$ and
$e_{ij} = 0$ for the other components of the strain tensor,
we obtain a stress tensor $\{ \tau_{ij} \}$ from which 
we calculate the stiffness constants $C_{13}$ and $C_{33}$.
Comparing the results of both methods provides us with
a consistency check for our calculations.

\section{Phonon dispersion bands and elastic constants at $T = 0$}

The evaluation of the elastic stiffness constants, $C_{ij},$ of graphite
with the LCBOPII potential model in the classical $T \rightarrow 0$ limit 
provides us with a useful reference for the subsequent analysis of 
temperature and nuclear quantum effects. 
Two alternative methods have been employed to derive $C_{ij}$ in this 
limit, namely the analysis of the harmonic dispersion relation of acoustic 
phonons, and the calculation of the elastic energy associated to some 
selected strain tensors, $\{ e_{ij} \}$. 

\begin{figure}
\vspace{0.2cm}
\includegraphics[width=7cm]{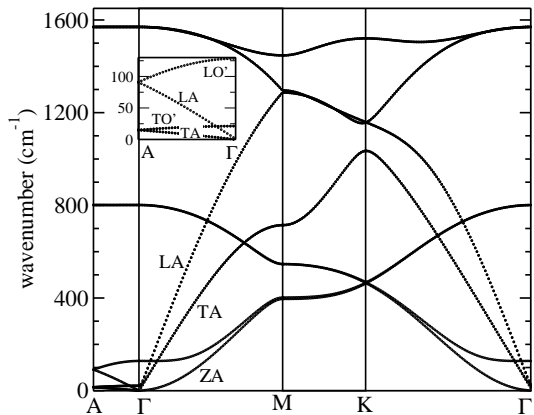}
\vspace{0.6cm}
\caption{Phonon dispersion bands of graphite as derived from
the LCBOPII potential model in a harmonic approximation.
The labels LA, TA, and ZA refer to the acoustic bands employed in
the calculation of the elastic stiffness constants.
The inset amplifies the low-energy acoustic bands along
the $\Gamma-A$ direction.
}
\label{f2}
\end{figure}

The phonon bands of graphite derived from the LCBOPII potential
employed here are displayed in Fig.~2.
These bands were obtained by diagonalization of the dynamical
matrix along selected symmetry directions in reciprocal space. 
Interatomic force constants were derived by numerical differentiation 
of the forces using atom displacements of $1.5\times10^{-3}$~\AA\ 
with respect to the equilibrium positions. 
The phonon dispersion in Fig.~2 is similar to those obtained from
other empirical potentials and DFT calculations.\cite{wi04}
In Table~I we present frequencies of optical modes at the $\Gamma$
point and acoustic modes at the high-symmetry points $M$, $K$, $A$
in $k$-space, derived from DFT calculations in the local-density
approximation (LDA) and generalized-gradient approximation (GGA),\cite{wi04}
as well as those obtained with the LCBOPII interatomic potential.

\begin{table}
\caption{Phonon frequencies (in cm$^{-1}$) at high-symmetry points in
$k$-space, obtained in Ref.~\onlinecite{wi04} from LDA and GGA-DFT
calculations, along with those found for the LCBOPII interatomic potential
employed here.
}
\vspace{0.6cm}
\centering
\setlength{\tabcolsep}{10pt}
\begin{tabular}{c c c c c}
          &            &    LDA   &     GGA   &   LCBOPII    \\[2mm]
\hline  \\[-3mm]
$\Gamma$  &   LO/TO    &  1597    &    1569   &    1570      \\
          &     ZO     &   893    &     884   &     901      \\
          &     TO'    &    43    &    ---    &      21      \\
          &     LO'    &   120    &    ---    &     128      \\
  $M$     &     LA     &  1346    &    1338   &    1266      \\
          &     TA     &   626    &     634   &     714      \\
          &     ZA     &   472    &     476   &     396      \\
  $K$     &   LA/LO    &  1238    &    1221   &    1158      \\
          &   ZA/ZO    &   535    &     539   &     465      \\
          &     TA     &  1002    &    1004   &    1035      \\
  $A$     &   LA/LO'   &    85    &    ---    &      90      \\
          &   TA/TO'   &    30    &    ---    &      14      \\[2mm]
\hline
\end{tabular}
\label{phonons}
\end{table}

The sound velocities for the three acoustic bands (LA, TA, and ZA) 
along the direction $\Gamma\mathrm{M}$,
with wavevectors $(k_{x},0,0)$, correspond to the slopes
$\left( \partial \omega / \partial k_x \right)_{\varGamma}$.
For the hexagonal symmetry of graphite, these velocities are 
related to the elastic stiffness constants as follows:\cite{ne05}
\begin{equation}
  \omega_{\rm LA}=\left(\frac{C_{11}}{\rho}\right)^{1/2}k_x  \:,
\end{equation}
\begin{equation}
  \omega_{\rm TA}=\left(\frac{C_{11}-C_{12}}{2\rho}\right)^{1/2}k_x  \:,
\end{equation}
\begin{equation}
  \omega_{\rm ZA}=\left(\frac{C_{44}}{\rho}\right)^{1/2}k_x  \:,
\end{equation}
where $\rho$ is the density of graphite. 
Along the $\Gamma\mathrm{A}$
direction, $(0,0,k_z)$, the sound velocities of the LA and the
twofold degenerate TA branches are given by,\cite{ne05}
\begin{equation}
  \omega_{\rm LA} = \left( \frac{C_{33}}{\rho} \right)^{1/2}k_z  \:,
\end{equation}
\begin{equation}
  \omega_{\rm TA} = \left( \frac{C_{44}}{\rho} \right)^{1/2}k_z  \:.
\end{equation}
The interatomic potential LCBOPII was employed earlier to
obtain the phonon dispersion bands of graphene.\cite{ka11}
We note that the version of the potential used in that work was
slightly different than that employed here, which is more 
realistic to describe the bending of the graphene 
sheets,\cite{ti17,ra16} as mentioned above in Sec.~II.

Our second approach to determine $C_{ij}$ at $T = 0$ consists in
calculating the elastic energy, $E_{\rm elas}$, 
for small strains $e_{ij}$, which can be expressed as
\begin{equation}
  \frac{E_{elas}}{V_0} = \frac{E-E_{0}}{V_{0}} =
      \frac12 \sum_{i=1}^6 \sum_{j=1}^6 C_{ij} e_i e_j  \:,
\end{equation}
where $E_0$ and $V_0$ are the energy and volume of the equilibrium
configuration in the absence of strain (see below).
We use the Voigt notation, where the components of the strain tensor
are labeled as
\begin{equation}
  \left\{ e_i, i = 1, \ldots 6 \right\} = 
   \left\{ e_{xx}, e_{yy}, e_{zz}, e_{yz}, e_{xz}, e_{xy} \right\} \:.
\end{equation}

\begin{table*}[ht]
\caption{Definition of the six different cell strains employed to
calculate the elastic energy and the classical value of the elastic
stiffness constants at $T = 0$.  The isotropic bulk modulus,
        $B_{iso}$, is defined in Eq.~(\ref{bulkm_iso}).}
\vspace{0.6cm}
\begingroup
\setlength{\tabcolsep}{10pt}
\renewcommand{\arraystretch}{1.5}
\begin{tabular}{|c|c|c|}
\hline
strain & components, $e_{ij} \neq 0$ & $E_{elas}/V_0$   \\ [2mm]
\hline
1 & $e_{xx}= e_{yy} = e$ &
    $(C_{11}+C_{12})e^{2}$   \\

2 & $e_{xx}=-e_{yy}= e$ &
    $(C_{11}-C_{12})e^{2}$  \\

3 & $e_{zz} = e$ &
    $\frac{1}{2}C_{33}e^{2}$  \\

4 & $e_{xz} = e_{zx} = e$ &
    $2C_{44}e^{2}$   \\

5 & $e_{xx} = e_{yy} = e_{zz} = e$$ $ &
    $\frac{9}{2}B_{iso}e^{2}$   \\

6 & $e_{xx} = e_{yy} = -\frac{e}{3}, \: e_{zz}=\frac{2 e}{3}$ &
    $\frac{1}{9}(C_{11}+C_{12}+2C_{33}-4C_{13}) e^2$  \\
\hline
\end{tabular}
\endgroup
\label{tab:strains}
\end{table*}

The elastic energy corresponding to six different strain tensors,
employed in the evaluation of $C_{ij}$, is summarized in Table~II.
The tensor components are defined with the help of a dimensionless
constant $e$. The elastic stiffness constants were obtained
by quadratic fits of the function $E_{elas}/V_{0}$ for strains defined
in the region $\left | e \right| \leq 2 \times 10^{-3}$. 
One important aspect
in the calculation of the elastic energy is that, whenever the solid
lattice is subjected to a uniform strain, the atoms will rearrange
themselves in the distorted lattice to minimize 
the elastic energy.\cite{fe77,st96,zh08}   This aspect is especially
important in the case of strain 2 in Table~II, where
upon a uniform distortion of the lattice, one finds that the elastic
energy is reduced by a 13\% when internal relaxation of atomic 
positions in the distorted lattice is allowed. 
Only when this atomic relaxation is included,
the elastic constants calculated by the elastic-energy method and
the acoustic phonon dispersion agree with each other.\cite{st96}
Our results for the classical $T \rightarrow 0$ limit of the elastic
stiffness constants of graphite are summarized in Table~III, along
with also finite-temperature values which will be
discussed below. The error bars in the classical zero-temperature 
values were derived from the differences encountered
between both methods employed here, except for $C_{13}$, where
the error bar corresponds to the results obtained 
with strains 5 and 6. 

\begin{table*}[ht]
\caption{Elastic stiffness constants, bulk modulus, and Poisson's
ratio of graphite, derived from classical and quantum simulations
at $T$ = 0, 300 and 750~K. Data for $C_{ij}$ and $B$ are in GPa.
We give in parenthesis the statistical
error in the last digit.
}
\vspace{0.6cm}
\begingroup
\setlength{\tabcolsep}{10pt}
\renewcommand{\arraystretch}{1.5}
\centering
\setlength{\tabcolsep}{10pt}
\begin{tabular}{c|c c| c c| c c|}
\cline{2-7}
\multicolumn{1}{c}{} &
      \multicolumn{2}{|c|}{$T = 0$} &
      \multicolumn{2}{c|}{$T = 300$ K} &
      \multicolumn{2}{c|}{$T = 750$ K} \\ [2mm]
\cline{2-7}
     & class.  & quantum & class. & quantum & class. & quantum
  \\[2mm]
\hline
 $C_{11}$ & 1007.7(5) & 992(1) & 969(1)& 960(2) & 917(2) & 910(2) \\

 $C_{12}$ & 216.3(3) & 174(1) & 175(1)& 162(2) & 134(1) & 131(2)  \\

 $C_{13}$ &  1.05(4) & 1.0(1) & 3.4(1) & 3.4(1)& 5.7(1)  & 5.7(1) \\

 $C_{33}$ &  37.1(1) & 35.9(1)& 34.6(1)& 33.9(1)& 31.6(1)& 31.2(1) \\

 $C_{44}$ &  1.03(2) & 0.74(1)& 0.61(1)& 0.57(1)& 0.42(2)& 0.41(2) \\

  $B$     &  35.1(1) & 33.8(1)& 33.0(1)& 32.3(1)& 30.4(1)& 30.0(1)  \\

  $\nu$   &   0.215  & 0.175  &  0.181 &  0.169 &  0.146 & 0.144    \\
\hline
\end{tabular}
\endgroup
\label{el_const_simul}
\end{table*}

\section{Energy}

In this section we present the internal energy of unstressed 
graphite, obtained from PIMD simulations in the isothermal-isobaric 
ensemble at various temperatures. This kind of simulations yield
separately the kinetic and potential energy of the 
system,\cite{he14,ra11} which allows us to analyze anharmonicities 
in the solid by comparing both energies.
For given temperature and external stress, we express 
the internal energy as $E = E_0 + E_{\rm pot} + E_{\rm kin}$,
where $E_{\rm pot}$ and $E_{\rm kin}$ are the kinetic and potential 
energy, and $E_0$ is the energy of the classical model at $T = 0$,
i.e., the minimum-energy configuration of the considered 
LCBOPII potential, with totally planar sheets and no atomic 
quantum delocalization.

In the classical minimum, the energy of graphite
decreases by 50 meV/atom with respect to an isolated graphene
monolayer. This stabilization energy, associated to 
layer interactions, is in line with that
found from classical Monte Carlo simulations of bilayer 
graphene using the LCBOPII potential\cite{za10b} 
(25 meV/atom, since in this case each graphene
layer has only one neighboring layer).  The interlayer 
binding energies obtained from various {\em ab-initio} 
calculations for the AB stacking of graphite display
a rather large dispersion, with most data between 
20 and 80 meV/atom.\cite{ha04,ha07,mo15b}
Experimental values at room temperature lie between
35 and 52 meV/atom.\cite{ha04}

\begin{figure}
\vspace{-0.6cm}
\includegraphics[width=7cm]{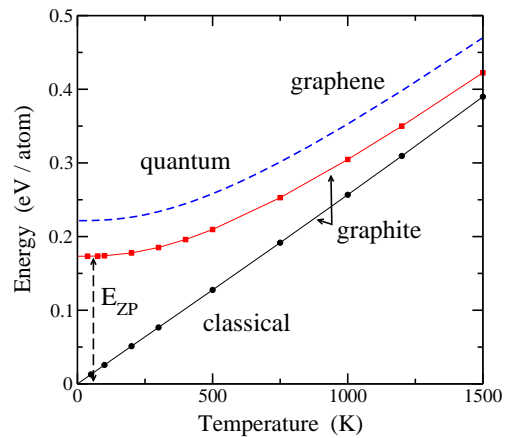}
\vspace{-0.5cm}
\caption{Internal energy of unstressed graphite as a function of
temperature, obtained from PIMD (squares) and classical MD
simulations (circles). Solid lines are guides to the eye.
Error bars are less than the symbol size. A vertical dashed arrow
indicates the zero-point energy, $E_{\rm ZP}$ = 173 meV/atom.
A dashed line shows the internal energy of monolayer graphene,
as derived from PIMD simulations.
}
\label{f3}
\end{figure}

In Fig.~3 we display the temperature dependence of the internal energy
per atom, $E - E_0$, as derived from our PIMD simulations of unstressed
graphite (solid squares). 
For comparison, we also show the internal energy obtained from
classical MD simulations (circles).
In the quantum case, $E - E_0$ converges in the low-$T$ limit to a
zero-point energy $E_{\rm ZP}$ = 173 meV/atom.
This value is slightly higher than that corresponding to 
a graphene monolayer\cite{he16} ($E_{\rm ZP} = 171$ meV/atom), 
which indicates that most of the zero-point energy is due to 
high-frequency in-plane vibrational modes, which are not 
appreciably changed by the interaction between layers.

In the quantum model, the internal energy follows at low 
temperature ($T < 200$ K) a dependence $E - E_0 \sim T^3$,
which is consistent with the known dependence
$c_p \sim T^2$ for the specific heat of graphite in this 
temperature region.\cite{de58,he20}
The classical model yields at low $T$ a dependence
$E - E_0 \sim T$, as expected from the equipartition principle
in classical statistical mechanics for harmonic lattice vibrations, 
which gives the Dulong-Petit law: 
$c_p = 3 N k_B$ irrespective of $T$.
At high $T$ we find from the classical simulations slight 
deviations from this law, due to anharmonicity of 
the vibrational modes.
The energy data found from PIMD simulations converge to those 
of classical MD simulations as temperature is raised. However at 
$T = 1000$~K we still observe a significant difference between 
quantum and classical results, close to 50~meV/atom.

The dashed line in Fig.~3 represents the internal energy per atom 
for a graphene monolayer, as derived from PIMD simulations.
This line is shifted upwards by $\Delta E =$~48 meV/atom with 
respect to the quantum data of graphite, 
which is the effective interlayer interaction.
This value is slightly lower than that found in the classical
calculation at $T = 0$ ($\Delta E =$~50 meV/atom), due to 
the larger zero-point energy per atom in graphite.

As indicated above, an overall quantification of the anharmonicity
of vibrational modes in graphite can be obtained by comparing
the kinetic and potential energy yielded by PIMD simulations.
For strictly harmonic vibrations, one has $E_{\rm kin} = E_{\rm pot}$
(virial theorem), so departure from pure harmonicity 
can be assessed in view of deviations from unity of the ratio 
$E_{\rm kin} / E_{\rm pot}$.
At low $T$, we find for graphite a ratio 
$E_{\rm kin} / E_{\rm pot} = 1.02$.
From earlier analysis of anharmonicity in solids, on the basis
of quasiharmonic approximations and perturbation theory,
it is known that for small $T$, changes in the vibrational energy
with respect to a harmonic approach are essentially due to 
the kinetic-energy contribution.
Indeed, for a perturbed harmonic oscillator with an energy
perturbation proportional to $r^3$ or $r^4$ at $T = 0$
(here $r$ is any coordinate in the problem), the first-order 
change in the energy is due to a variation of $E_{\rm kin}$, and
$E_{\rm pot}$ keeps changeless as in the unperturbed
oscillator.\cite{he95,la65}

\section{Structural variables}

\subsection{Crystal volume}

\begin{figure}
\vspace{-0.6cm}
\includegraphics[width=7cm]{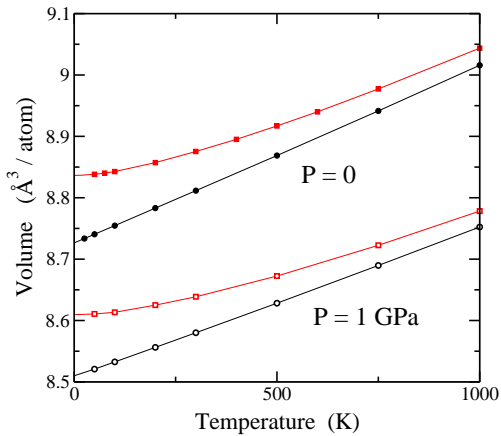}
\vspace{-0.5cm}
\caption{Temperature dependence of the crystal volume of
graphite, as derived from PIMD (squares) and classical MD (circles)
simulations for $P = 0$ (solid symbols) and a hydrostatic
pressure $P = 1$~GPa (open symbols).
Error bars are smaller than the symbol size.
}
\label{f4}
\end{figure}

In Fig.~4 we present the volume per atom as a function of
temperature, as derived from classical MD (circles) and PIMD
simulations (squares) in the $NPT$ ensemble for zero external
stress (solid symbols) and a hydrostatic pressure $P = 1$~GPa
(open symbols). As is usual in Thermodynamics, and in the
definition of the bulk modulus considered below, a compressive
pressure is positive.
In the stress-tensor notation employed in elasticity
this corresponds to $\tau_{xx} = \tau_{yy} = \tau_{zz} = -1$~GPa,
and $\tau_{\alpha \beta} = 0$ for $\alpha \neq \beta$.

We comment first the results for the unstressed material.
The data derived from classical simulations display a temperature
dependence of the volume close to linear, with a positive slope
$d V/ d T$ slowly increasing for rising $T$.
These data converge at low temperature to 
$V_0 = 8.727$~\AA$^3$/atom, which corresponds to the
minimum-energy volume.  For comparison, we mention that earlier
theoretical work based mainly on DFT calculations 
gave values for $V_0$ between 8.61 and 
8.94~\AA$^3$/atom.\cite{ja87,sc92,fu94,bo97}
The equilibrium volume obtained from PIMD simulations at each
temperature is larger than the classical result, and 
converges to 8.837~\AA$^3$/atom for $T \to 0$. 
This represents a zero-point volume expansion of 1.3\% with 
respect to the classical minimum.
At high temperature the quantum and classical data converge
one to the other.
Our results are not far from the volume obtained for graphite
at ambient conditions from x-ray diffraction experiments:
8.78 and 8.80~\AA$^3$/atom in Refs.~\onlinecite{ha89}
and \onlinecite{zh89}, respectively.

At low temperature ($T \to 0$) the interlayer distance, $c$,
increases from the classical limit by a 0.4\%,
and the linear expansion in the sheet plane ($x$ and $y$
directions) amounts also to a 0.4\%. 
The zero-point expansion of a crystal is controlled by the 
anharmonicity of the lattice vibrations. In terms of a QHA, 
this expansion depends on each phonon through the product of 
its zero-point energy and the corresponding Gr\"uneisen 
parameter.\cite{as76,de96,mo05,he20c}
Given that the main contributions to the in-plane
and out-of-plane expansions are dominated by phonons with
different polarization, 
it seems accidental that the relative quantum effects in 
directions $x$ and $z$ coincide, in spite of the large
anisotropy of the material. 
This anisotropy is, however, clearly
observable in the thermal expansion at finite temperatures 
(see Sec.~VI).

For graphite under a hydrostatic pressure $P = 1$~GPa,
we obtain classical and quantum results similar to
those found for the unstressed material, with the following
differences. First, the volume is reduced, but this reduction
is much less in the $(x, y)$ plane than in the out-of-plane
$z$-direction, which corresponds to the different magnitudes
of the elastic constants governing the compressibility in 
different crystal directions (see below).
Second, the zero-point volume expansion is reduced with respect
to that found for unstressed graphite. For $P = 1$~GPa we
find $\Delta V$ = 0.099 \AA$^3$/atom for $T \to 0$ vs an
expansion of 0.110 \AA$^3$/atom for $P = 0$.
Third, the thermal dilation decreases under an applied
hydrostatic pressure. In the temperature range from $T$ = 0
to 1000~K, we find from the PIMD data an expansion 
of 0.169 and 0.207 \AA$^3$/atom for the stressed and unstressed
material, respectively.

\begin{figure}
\vspace{-0.6cm}
\includegraphics[width=7cm]{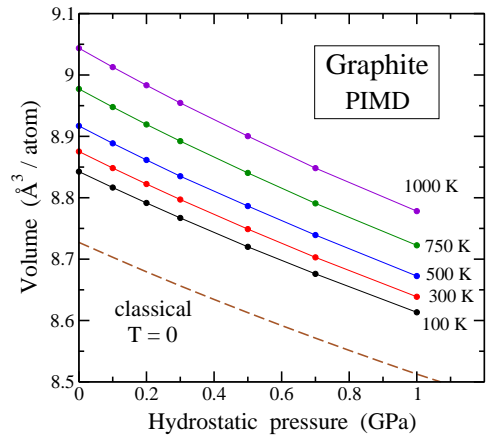}
\vspace{-0.5cm}
\caption{Volume of graphite as a function of hydrostatic
pressure at various temperatures, as derived from PIMD
simulations. From top to bottom, $T$ = 1000, 750, 500,
300, and 100~K. Solid lines are guides to the eye.
A dashed line represents the pressure dependence of
the classical volume at $T = 0$.
}
\label{f5}
\end{figure}

The dependence of crystal volume on hydrostatic pressure is
displayed in Fig.~5 for several temperatures.
Symbols are data points derived from PIMD simulations and
solid lines are guides to the eye.
The dashed line indicates the pressure dependence of $V$
for the classical model at $T = 0$. For each pressure
$P$, this corresponds to the minimum of the enthalpy
$H = E - P V$.
Note the important difference between this classical result 
at $T = 0$ and the quantum result at $T = 100$~K 
(more than 0.1~\AA$^3$/atom), due to the low-temperature 
crystal expansion for the quantum model, as shown in Fig.~4.
The solid lines in Fig.~5 seem at first sight rather parallel, 
but differences between their slopes (in particular at $P = 0$) 
indicate changes in the bulk modulus of the material (see below).
The volume differences between these isotherms become
smaller as the hydrostatic pressure is increased.
Thus, the difference between the quantum result at 100~K and
the classical minimum is reduced by a 13\% from $P$ = 0
to 1~GPa.

Volume changes under a hydrostatic pressure are related to
the bulk modulus $B$, as discussed in Sec.~VIII.
From the classical data at $T = 0$, we have
$d V / d P = -0.246$~\AA$^3$/(atom GPa) in the limit $P \to 0$, 
whereas at $T$ = 1000~K, our PIMD simulations yield
$d V / d P = -0.313$~\AA$^3$/(atom GPa).

The strain components $e_{xx}$ and $e_{zz}$ for
a hydrostatic pressure $P$ can be obtained from the elastic 
compliance constants $S_{ij}$ as\cite{ma18,ra20b}
\begin{equation}
 \frac{e_{xx}}{e_{zz}} = \frac{S_{11} + S_{12} + S_{13}}
          {2 S_{13} + S_{33}}   \; .
\end{equation}
At room temperature ($T = 300$~K) we find from our PIMD simulations
$e_{xx} / e_{zz}$ = 0.027. At 1000~K this ratio decreases to 0.023.

\begin{figure}
\vspace{-0.6cm}
\includegraphics[width=7cm]{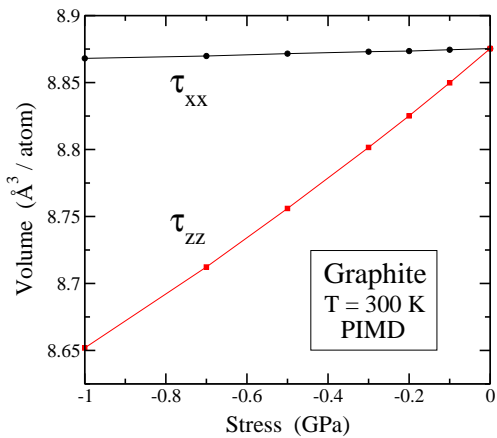}
\vspace{-0.5cm}
\caption{Volume of graphite as a function of uniaxial stress
along the $x$ and $z$ directions of the orthorhombic simulation
cell at $T$ = 300~K.  $\tau_{xx}$ and $\tau_{zz}$ indicate
the nonvanishing components of the stress tensor in each case.
Symbols represent results of PIMD simulations,
with error bars less than the symbol size.
Lines are guides to the eye.
}
\label{f6}
\end{figure}

For the calculation of elastic constants at finite temperatures 
presented in Sec.~VII, 
we have carried out simulations of graphite under various kinds of 
stress, given by the different components of the tensor 
$\{ \tau_{ij} \}$, as explained in Sec.~II.
In Fig.~6 we present the dependence of the crystal volume on
uniaxial stress along the $x$ and $z$ directions at $T$ = 300~K,
as derived from PIMD simulations. 
These uniaxial stresses correspond to nonvanishing components 
of the stress tensor $\tau_{xx}$ and $\tau_{zz}$, respectively. 
The volume change in the second case is much
larger than in the former, due to the higher compressibility
in the $z$ direction, perpendicular to the layer planes.
For $\tau_{xx}$ and $\tau_{zz}$ close to zero we find
for the stress derivative of the volume values of
$-7.2 \times 10^{-3}$ and -0.256 \AA$^3$/(atom GPa), respectively.

The volume change under a uniaxial stress can be obtained
from the elastic constants of the material, in particular
from the compliance constants. We have\cite{ma18,ra20b} 
\begin{equation}
 \frac{\Delta V}{V} = e_{xx} + e_{yy} + e_{zz} =
            (S_{11} + S_{12} + S_{13}) \tau_{xx}  \; ,
\end{equation}
or, for the stress derivative of the volume:
\begin{equation}
 \frac{d V}{d \tau_{xx}} = (S_{11} + S_{12} + S_{13}) V  \; .
\end{equation}
Similarly, for a uniaxial stress $\tau_{zz}$, one has
\begin{equation}
 \frac{d V}{d \tau_{zz}} = (2 S_{13} + S_{33}) V  \; .
\end{equation}
Note that the ratio of these stress derivatives of the volume
coincides with the ratio $e_{xx} / e_{zz}$ given above for
a hydrostatic pressure.

\subsection{In-plane area}

\begin{figure}
\vspace{-0.6cm}
\includegraphics[width=7cm]{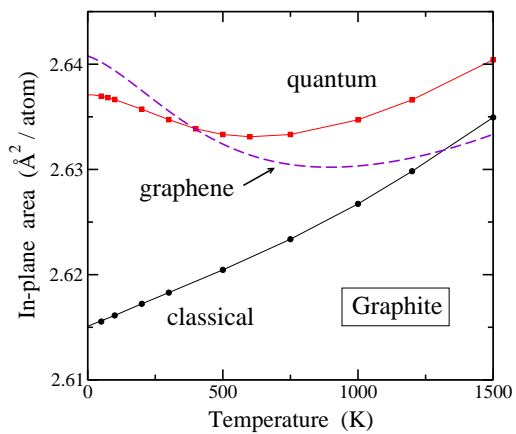}
\vspace{-0.5cm}
\caption{Temperature dependence of the in-plane area per atom,
$A_p$, obtained from classical (circles) and PIMD (squares)
simulations for $n = 240$.
Error bars are less than the symbol size.
Solid lines are guides to the eye.
A dashed line indicates the dependence of $A_p$ on $T$ for
a graphene monolayer, as derived from PIMD simulations for the
same system size.
}
\label{f7}
\end{figure}

In the graphene literature, it has been discussed with great
detail the behavior of the in-plane area of the 2D material,
which in the case of graphite corresponds to the area $L_x L_y$ 
on the $xy$ plane of the simulation box. Here we will consider 
the in-plane area per atom, $A_p = L_x L_y / n$.
In Fig.~7 we show the temperature dependence of $A_p$ for
unstressed graphite, as derived from 
classical MD (circles) and PIMD simulations (squares).
At first sight one observes an important difference between
the quantum and classical results.
In the quantum data we find a decrease in $A_p$ for rising
$T$ until a temperature of about 600~K, for which it reaches
a minimum, and an increase in $A_p$ at higher $T$.
In contrast, in the classical data we obtain a rise 
(roughly linear) of $A_p$ at low $T$, and an increase faster 
than linear at $T > 500$~K. 
At $T = 1500$~K the difference between classical and quantum
data for $A_p$ is still much larger than the error bars of the
simulation results (smaller than the symbol size in Fig.~7).

In the low-temperature limit ($T \to 0$), the 
in-plane area $A_p$ derived from PIMD simulations converges to
2.6371 \AA$^2$/atom, with a C--C bond length 
$d_{\rm C-C} =$ 1.4276~\AA.
For the classical minimum we find a value $A_0 = 2.6149$ \AA$^2$/atom,
which corresponds to a C--C distance $d_0 = 1.4188$~\AA.
This gives for the quantum result a zero-point expansion in 
the area $A_p$ of 0.022 \AA$^2$/atom, i.e. a relative increase 
of about 1\%, associated to the rise in C--C bond length.
Looking at the C--C distance for the classical minimum, $d_0$, 
we observe that there is
a slight in-plane lattice contraction with respect to the
cases of a graphene monolayer ($d_0$ = 1.4199~\AA) and
bilayer ($d_0$ = 1.4193~\AA), as a consequence of interlayer
interactions.\cite{he16,he19}
We note that, although in the classical model one has
planar carbon sheets for $T \to 0$, in the quantum zero-temperature 
limit the layers are not exactly planar, since there is an 
atomic zero-point motion in the out-of-plane 
$z$-direction.\cite{he16,ra17} 

For comparison with the results for graphite, we also present
in Fig.~7 the area $A_p$ obtained from PIMD simulations of
monolayer graphene with the same size $n = 240$ (dashed line). 
In this case, the shape of the
temperature dependence is similar to that of graphite,
with a decrease in $A_p$ at low $T$ and an increase at high $T$.
For graphene, however, the decrease is larger and
the minimum of $A_p$ occurs at a higher temperature.
At low $T$, the area $A_p$ of graphite is reduced by 
$4 \times 10^{-3}$ \AA$^2$/atom with respect to a graphene
monolayer, as a consequence of layer interactions.

The fact that $d A_p / d T < 0$ at low temperature, as obtained
in the quantum simulations, is due to the out-of-plane motion of the
carbon atoms, which dominates over the thermal expansion of
the C--C bonds in the graphite layers.
This effect is not captured by a classical model for the atomic
motion, as happens in classical MD simulations, where the
relative contributions of the different vibrational modes
are not correctly represented at low temperatures.  
At high $T$, the bond expansion dominates over the contraction
associated to motion in the $z$-direction, so that
$d A_p / d T > 0$ in both classical and quantum models.

\section{Thermal expansion}

At low $T$ our PIMD simulations give for graphite with AB
stacking an interlayer spacing $c$ = 3.3510~\AA.
For the classical model at $T = 0$ (minimum energy configuration
with planar graphene sheets) we find $c_0$ = 3.3372~\AA.
Thus, we have a zero-point expansion 
$\Delta c = 1.4 \times 10^{-2}$~\AA.
At $T$ = 300 K, we have $c$ = 3.3688~\AA\ from
PIMD simulations, not far from a distance of 3.3538 \AA\ obtained 
by Baskin and Meyer from x-ray diffraction measurements
at room temperature ($T = 297$~K).\cite{ba55} 
At 300 K, the difference between
quantum and classical results is around four times smaller
than for the low-temperature limit.

\begin{figure}
\vspace{-0.6cm}
\includegraphics[width=7cm]{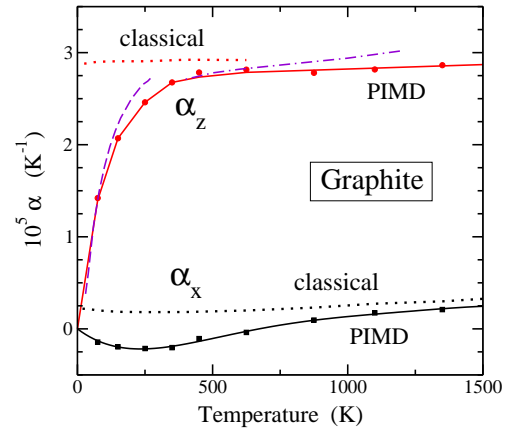}
\vspace{-0.5cm}
\caption{Linear thermal expansion coefficients of graphite vs
temperature: $\alpha_x$ along the in-plane $x$-direction and
$\alpha_z$ along the out-of-plane $z$-direction.
Symbols represent data obtained from PIMD simulations for
$n = 960$ and solid lines are guides to the eye.
Dotted lines indicate results derived from classical MD simulations.
The dashed line represents data obtained for $\alpha_z$ by
Bailey and Yates\cite{ba70} from interferometric measurements
at $T < 300$~K. The dashed-dotted line is a fit to experimental
data of $\alpha_z$ for $T > 300$~K.\cite{ma18}
}
\label{f8}
\end{figure}

From the mean interlayer spacing we consider the linear
thermal expansion coefficient (TEC) $\alpha_z$, defined as
\begin{equation}
 \alpha_z = \frac{1}{c}
      \left( \frac{\partial c}{\partial T} \right)_{\tau}   \, .
\end{equation}
This TEC for vanishing external stress has been commonly denoted as 
$\alpha_c$ in the literature, but we will call it here $\alpha_z$ 
for notation consistency.
Data for $\alpha_z$ obtained from PIMD simulations of 
graphite are presented in Fig.~8 as solid circles. 
These data points were obtained from a numerical derivative
of the mean layer spacing $c$ found in the simulations at
several temperatures.
One observes a fast rise of $\alpha_z$ in the low-temperature
region up to around 200 K, and at higher $T$ this rise becomes 
much slower. At 300~K we find 
$\alpha_z = 2.6 \times 10^{-5}$~K$^{-1}$. 
For comparison, we also present in Fig.~8 results of classical MD 
simulations for $\alpha_z$ (dotted line). They converge 
at low $T$ to a value $\alpha_z = 2.9 \times 10^{-5}$~K$^{-1}$.
Note the inconsistency of this classical result with the third
law of Thermodynamics, which requires that TECs should vanish
for $T \to 0$.\cite{ca60,as76}

Experimental data for $\alpha_z$ of pyrolytic graphite at low $T$ 
were obtained by Bailey and Yates\cite{ba70} from interferometric
measurements (dashed line in Fig.~8).
The dashed-dotted line indicates a fit to experimental data
from several sources at $T > 300$~K, presented by
Marsden {\em et al.}\cite{ma18}
Both dashed and dashed-dotted lines derived from experimental 
data do not fit well one with the other close to room temperature, 
due to the dispersion of data in different source references. 
At $T > 500$~K we observe that the TEC $\alpha_z$ obtained
from our PIMD simulations rises slower than the line fitted to
experimental data in Ref.~\onlinecite{ma18}.

In the $xy$ layer plane, we consider a linear TEC defined as
\begin{equation}
 \alpha_x = \frac{1}{L_x}
      \left( \frac{\partial L_x}{\partial T} \right)_{\tau}   \, .
\end{equation}
In Fig.~8 (bottom) we display $\alpha_x$ obtained from PIMD 
simulations of graphite up to 1500 K (solid squares).
The solid line represents a polynomial fit to the data points.
At low temperature this TEC is negative and reaches a minimum at
$T_m \approx 250$~K. For $T > T_m$, $\alpha_x$ increases for
rising temperature and becomes positive at $T_m \approx 600$~K,
which coincides with the temperature at which the in-plane area 
$A_p$ attains its minimum value, as shown in Fig.~7.
For comparison, the dotted line in Fig.~8 (bottom) represents
the results obtained for $\alpha_x$ from classical MD simulations.
This classical $\alpha_x$ takes positive values
in the whole temperature region considered here,
and converges at low $T$ to a (nonphysical) value of 
$2.2 \times 10^{-6}$~K$^{-1}$. 
The quantum data for $\alpha_x$ are below the classical ones,
and for $T > 1500$~K they are close one to the other.

Our results for $\alpha_x$ show a temperature dependence 
similar to those obtained earlier employing other theoretical
techniques, in particular that found
by Mounet and Marzari\cite{mo05} from a combination of
DFT calculations with a QHA for the vibrational modes.  These
authors found for $\alpha_x$ a minimum of 
$-1.2 \times 10^{-6}$~K$^{-1}$ for $T \approx 250$~K
and a vanishing TEC for $T \approx 500$~K.
Experimental results for the TEC $\alpha_x$ show a minimum  at a
temperature between 200 and 300 K, similarly to the data obtained 
from our quantum simulations.\cite{ke64,mo72,ma18} 
Several experimental data sets display a minimum of about 
$-1.5 \times 10^{-6}$~K$^{-1}$, which turns out to be
somewhat less than our data presented in Fig.~8.

\section{Linear elastic constants at finite temperatures}

\begin{figure}
\vspace{-0.6cm}
\includegraphics[width=7cm]{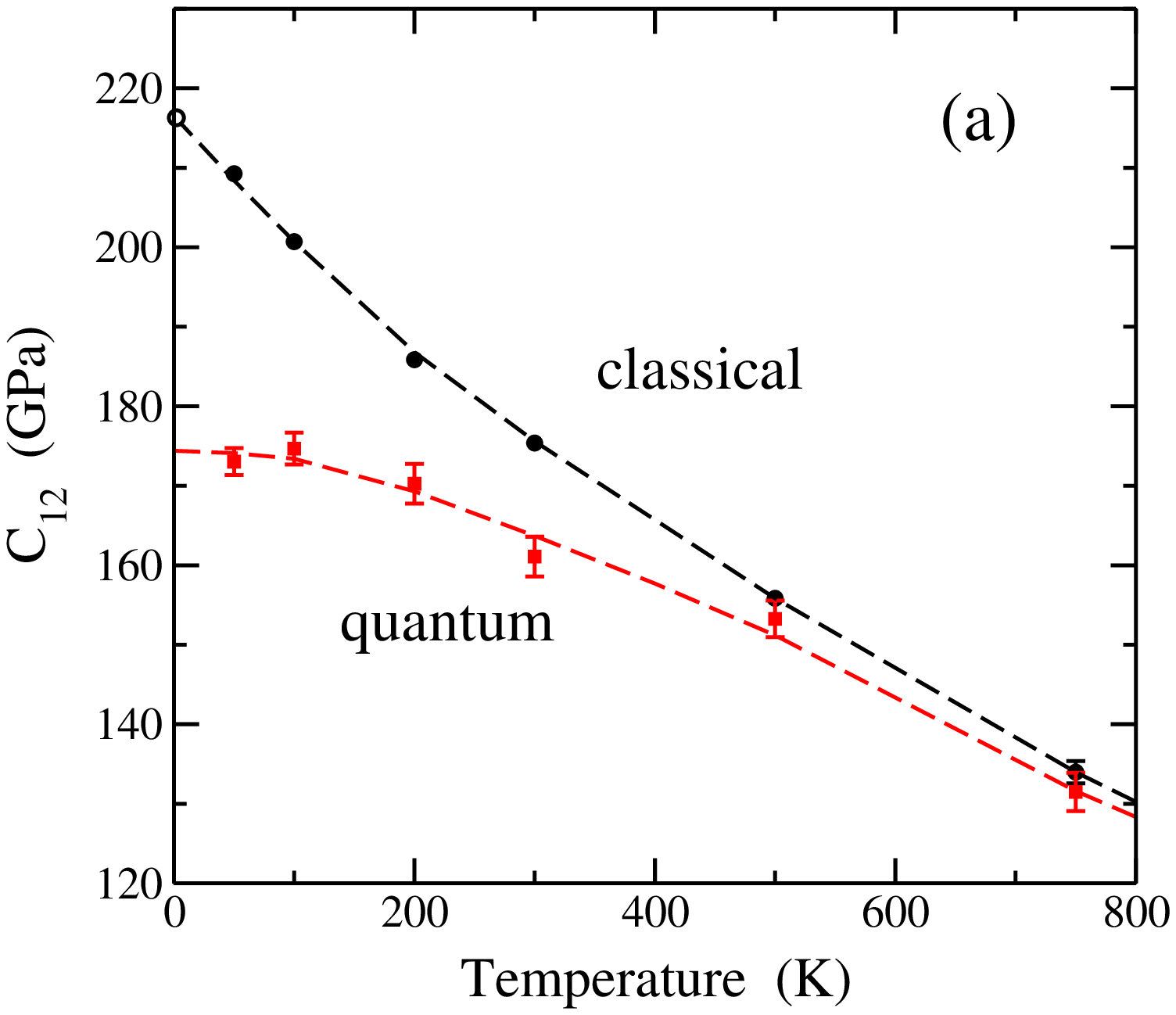}
\includegraphics[width=7cm]{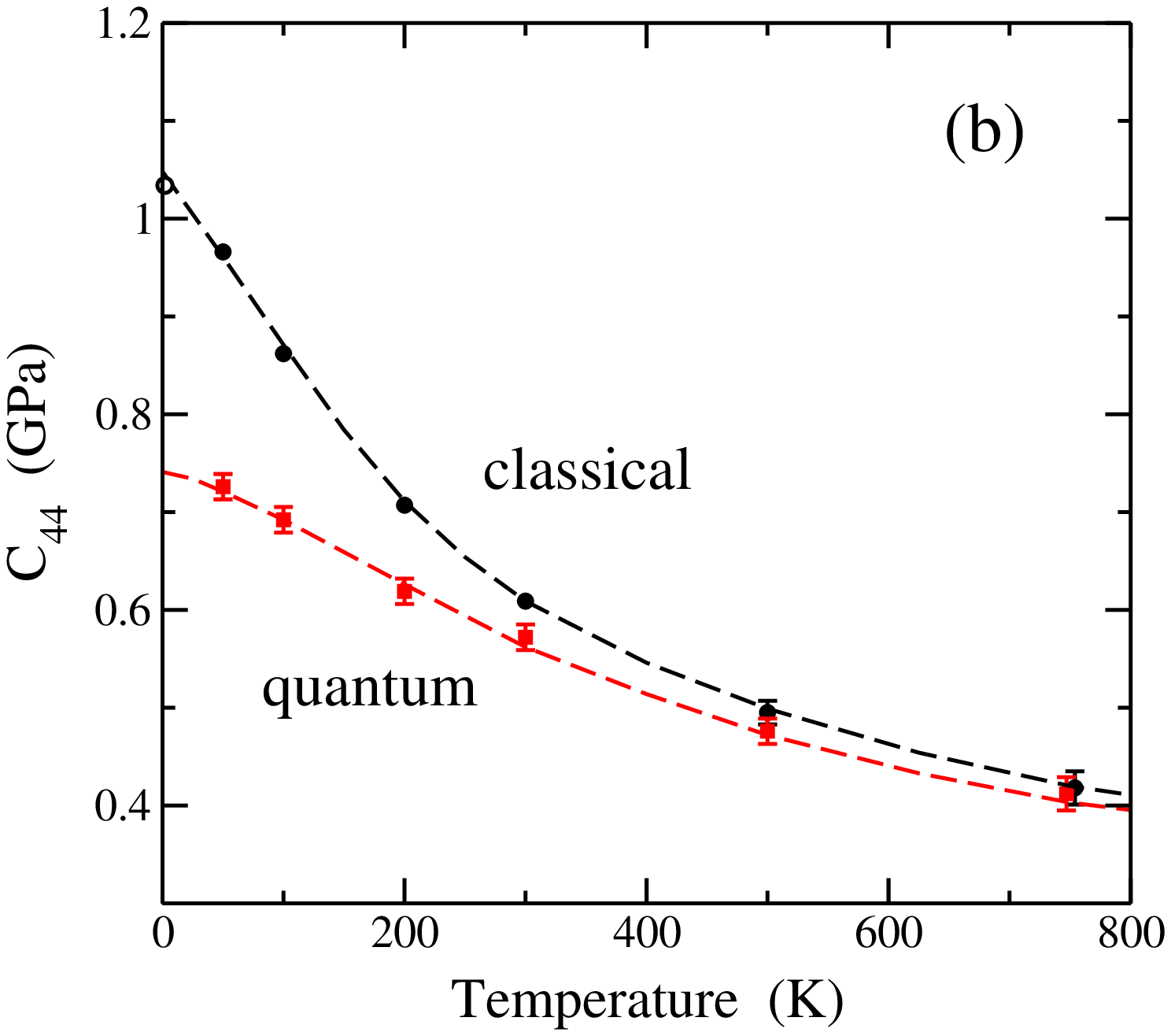}
\vspace{-0.5cm}
\caption{Temperature dependence of the elastic constants:
(a) $C_{12}$ and (b) $C_{44}$, as derived from classical MD (circles)
and PIMD (squares) simulations in the $NVT$ ensemble.
Open circles in (a) and (b) represent the classical value at $T = 0$,
calculated from the phonon dispersion curves.
Error bars, when not shown, are in the order of the symbol size.
Lines are guides to the eye.
}
\label{f9}
\end{figure}

In this section we present and discuss nuclear quantum effects
in the elastic stiffness constants of graphite. 
Such effects are present in general for the different 
elastic constants, mainly at low temperature, but they turn
out to be especially large for $C_{12}$ and $C_{44}$.
In Fig.~9(a) we present the temperature dependence of the
elastic constant $C_{12}$, as derived from our 
classical (circles) and PIMD (squares) simulations.
The classical finite-temperature results 
converge at low $T$ to the value obtained from the phonon
dispersion, indicated by an open circle in Fig.~9(a).
In the limit $T \to 0$, this elastic constant
is found to decrease from the classical value of 216 GPa
to 174 GPa due to zero-point motion. This represents a
reduction of a 19\% with respect to the classical result.
At room temperature the quantum reduction amounts to a 7\%.

In Fig.~9(b) we display the dependence of $C_{44}$ on temperature 
for both quantum (squares) and classical (circles) cases.
The open circle at $T = 0$ represents the value calculated from
the phonon dispersion curves as explained in Sec.~III
($C_{44} = 1.03(2)$~GPa).
The quantum results converge at low $T$ to $C_{44} = 0.74$~GPa.
This elastic constant is particularly interesting from the viewpoint
of quantum effects. Given its small value in comparison to
other stiffness constants of graphite, at low temperature the quantum 
correction with respect to the classical result is very large.
In the limit $T \to 0$, it means a relative reduction of 
$C_{44}$ by a 28\%.

Concerning nuclear quantum effects,
something similar occurs for other elastic stiffness constants,
as $C_{11}$ and $C_{33}$, for which results of
classical and PIMD simulations are given in Table~III at 
$T =$~300 and 750~K, as well as for the low-$T$ limit.
In all cases, the classical value at $T = 0$ is calculated from
the phonon dispersion bands and lattice strains, as explained
in Sec.~III. The low-temperature quantum values are obtained 
from an extrapolation of finite-temperature PIMD results.
For $C_{11}$ and $C_{33}$, zero-point motion causes a decrease
of 1.5\% and 3.0\% with respect to the classical value, respectively.
We do not clearly observe any quantum effect in $C_{13}$. 
In fact, for this stiffness constant the results of PIMD and 
classical MD simulations coincide within error bars in the 
temperature region considered here. 

It is worthwhile commenting on the relation between 
the temperature dependence of the linear elastic constants 
shown in Fig.~9 and the quantum delocalization of atomic
nuclei. For $C_{12}$ and $C_{44}$, we find in the classical 
results a decrease as temperature is raised. 
This is related to classical thermal
motion of the carbon atoms, which grows with temperature as
indicated by the MSD shown in Fig.~1. In the results of our
quantum PIMD simulations we observe an important decrease in the
zero-temperature elastic constants, due to zero-point
delocalization (finite MSD), with respect to the classical
values, where the atomic MSD vanishes.
As temperatures increases, the quantum and classical results
converge one to the other, as happens for the MSD.
Something similar can be said for the results of the bulk
modulus presented below in Sec.~VIII.

\begin{table}
\caption{Elastic stiffness constants, bulk modulus $B$, and
Poisson's ratio $\nu$ of graphite obtained from
various experimental techniques by different authors at ambient
conditions. Data for $C_{ij}$ and $B$ are given in GPa.
The bulk modulus $B$ is obtained in each case from
the elastic constants by using Eq.~(\ref{bulkm2}).
}
\vspace{0.6cm}
\centering
\setlength{\tabcolsep}{10pt}
\begin{tabular}{c c c c}
     & Blakslee\cite{bl70} &  Bosak\cite{bo07b} & Nicklow\cite{ni72}
  \\[2mm]
\hline  \\[-2mm]
  $C_{11}$  & 1060(20)  & 1109(16) & 1440(200)    \\[2mm]

  $C_{12}$  & 180(20)  & 139(36)   &  ---         \\[2mm]

  $C_{13}$  & 15(5)    &   0(3)    &  ---         \\[2mm]

  $C_{33}$  & 36.5(1)  & 38.7(7)   & 37.1(5)      \\[2mm]

  $C_{44}$  &  0.27(9) & 5.0(3)    & 4.6(2)       \\[2mm]

    $B$     & 35.8(2)  &  36.4(11) &   ---        \\[2mm]

    $\nu$   &  0.17    &   0.13    &   ---        \\[2mm]

\hline  \\[-2mm]
\end{tabular}
\label{el_const_exp}
\end{table}

In Table~IV we present values of the elastic stiffness constants
derived from experimental data by several authors,
from a combination of ultrasonic, sonic resonance, and static 
test methods,\cite{bl70} 
as well as inelastic x-ray scattering,\cite{bo07b} and
inelastic neutron scattering along with a force model.\cite{ni72}
There appears some dispersion in these results derived
from experiments at ambient conditions, in particular
for $C_{13}$ and $C_{44}$, as can be seen in our Table~IV
and in Refs.~\onlinecite{bl70,bo07b}.
For $C_{44}$ these values range from 5.0~GPa to less than 1~GPa. 
The low value $C_{44} = 0.27(9)$~GPa obtained by 
Blakslee {\em et al.}\cite{bl70} may be due to the presence 
of dislocations in the studied material, as suggested by 
the authors.
In a later paper, Seldin and Nezbeda\cite{se70} found that
this elastic constant rises when the graphite samples are
irradiated with neutrons at several temperatures.
These authors found that natural graphite crystals have after
irradiation a shear modulus $C_{44}$ in the range 1.6--4.6~GPa. 
Experimental data for $C_{13}$ of graphite are scarce, and
different techniques have yielded diverse outcomes. 
The results found by Bosak {\em et al.}\cite{bo07b} were 
compatible with a vanishing $C_{13}$ (within their error bar). 

In Table~V we give values of the elastic constants of graphite
calculated by various research groups.
Several calculations were carried out in the framework
of DFT, with both local-density approximation
(LDA) and generalized-gradient approximation (GGA).\cite{bo97,ha04,mo05}
Moreover, Jansen and Freeman\cite{ja87} employed the full-potential 
linearized augmented-plane wave (FLAPW) method, and 
Michel and Verberck obtained the elastic constants from the phonon 
spectrum calculated with an effective potential.\cite{mi08b}
In spite of the general reliability of these theoretical procedures,
there are some discrepancies between the results of the different 
research groups.
A common feature of the data derived from DFT calculations is that 
they yielded $C_{13} < 0$, as shown in Table~V.
Although this is not forbidden for the stability
of hexagonal crystals,\cite{mo14}
we are not aware of any experimental work on graphite where 
$C_{13}$ was found to be negative.
We also note the anomalous (very small) value  obtained for $C_{33}$ 
in Ref.~\onlinecite{ha04} from DFT-GGA calculations, which seems to
be due to a largely underestimated interlayer interaction.
Our main conclusion concerning the linear elastic constants of graphite is
that the intrinsic difficulty of calculating the elastic constants
of this largely anisotropic material, is still more complicated
at temperatures lower than the Debye temperature of the material,
where nuclear quantum effects are relevant.

\begin{table*}
\caption{Elastic stiffness constants, bulk modulus, and Poisson's
ratio of graphite obtained from
various calculations based on density-functional theory with LDA
and GGA, as well as FLAPW and lattice dynamics (Latt. dyn.).
Values indicated with an asterisk ($^*$) correspond to
$C_{11} + C_{12}$.  Data for $C_{ij}$ and $B$ are given in GPa.
The bulk modulus $B$ is obtained in each case from
the elastic constants by means of Eq.~(\ref{bulkm2}).
}
\vspace{0.6cm}
\centering
\setlength{\tabcolsep}{10pt}
\begin{tabular}{c c c c c c}
    & LDA\cite{bo97} & FLAPW\cite{ja87} & LDA, GGA\cite{ha04} & LDA, GGA\cite{mo05} & Latt. dyn.\cite{mi08b}
  \\[2mm]
\hline  \\[-2mm]
  $C_{11}$  &1279.6$^*$ & 1430$^*$ &  ---      & 1118, 1079   & 1211.3 \\[2mm]

  $C_{12}$  & ---      &   ---     &  ---      &  235, 217    & 275.5  \\[2mm]

  $C_{13}$  &  --0.5   &   --12    &  ---      & --2.8, --0.5 &  0.59 \\[2mm]

  $C_{33}$  &  40.8    &   56      & 30.4, 0.8 & 29.5, 42.2   & 36.79 \\[2mm]

  $C_{44}$  &  ---     &   ---     &  ---      &  4.5, 3.9    & 4.18  \\[2mm]

    $B$     &  38.3    &   50.2    &   ---     &  28.0, 39.6  & 35.1  \\[2mm]

    $\nu$   &  ---     &   ---     &   ---     & 0.210, 0.201 & 0.227  \\[2mm]
\hline  \\[-2mm]
\end{tabular}
\label{el_const_calc}
\end{table*}

From the elastic stiffness constants one can obtain 
the Poisson's ratio $\nu$, which is a measure of the relation
between the transverse and longitudinal strains under
an applied stress.
For graphite one has $\nu = C_{12} / C_{11}$.
In Table~III we give the Poisson's ratio calculated from 
the elastic constants yielded by our classical and quantum 
simulations.  In our results, $\nu$ is found to decrease as 
temperature is raised.
In the low-temperature limit, nuclear quantum motion causes
a reduction of $\nu$ from 0.215 to 0.175, i.e., it decreases
by a 19\%. At $T$ = 300 K the classical and quantum values are
0.181 and 0.169, respectively, with a decrease of a 7\% due to
quantum motion. At 750 K this decrease is small, about 1\%.

In Table~IV we give values of the Poisson's ratio obtained from
the elastic constants found in experimental works.\cite{bl70,bo07b}
The value derived from the work of Blakslee {\em et al.}\cite{bl70}
is close to our quantum result at $T = 300$~K, and that from
the paper by Bosak {\em et al.}\cite{bo07b} is somewhat lower.
Data derived from theoretical methods given in
Table~V are close to our classical value at $T = 0$,
$\nu = 0.215$.  A larger collection of data for the Poisson's 
ratio of graphite derived 
from theoretical methods is given in Ref.~\onlinecite{po15}. 
The data reported in the literature display a large dispersion, 
most of them lying in the region from 0.12 to 0.3.
We note that the Poisson's ratio usually considered for
graphite is an in-plane variable, i.e., it refers to
$x$ and $y$ directions. One can equally define an out-of-plane
ratio $\nu_{xz}$, referring to $x$ and $z$ directions.
In this case, $\nu_{xz} = C_{13} / C_{11}$, and we find 
from our PIMD simulations at 300~K: 
$\nu_{xz} = 3.5 \times 10^{-3}$.

\section{Bulk modulus}

We present here results for the isothermal bulk modulus,
$B = - V (\partial P / \partial V)_T$ , 
derived from our classical MD and quantum PIMD simulations.
To check the overall consistency of our procedures, we
obtain $B$ in three different ways: (1) calculating
${\partial P} / {\partial V}$ from 
a numerical differentiation for positive and negative 
hydrostatic pressures $P$ close to $P = 0$,
(2) from the volume fluctuations along $NPT$ simulation runs
at temperature $T$, and (3) from the elastic constants obtained
from the simulations.

In the isothermal-isobaric ensemble (our second method), 
the bulk modulus can be directly calculated from the 
mean-square fluctuations of the volume, 
$\sigma_V^2$, using the formula\cite{la80,he08}
\begin{equation}
     B = \frac{k_B T V}{\sigma_V^2}   \; ,
\label{bulkm}
\end{equation}
$k_B$ being Boltzmann's constant.
This expression has been employed earlier to obtain the bulk 
modulus of various kinds of solids from path-integral
simulations.\cite{he01,he08}

The bulk modulus of graphite can be also calculated from the
elastic constants of the material (our third procedure).
For a hydrostatic pressure $P$, we have
$\tau_{xx} = \tau_{yy} = \tau_{zz} = -P$, 
so that
\begin{equation}
 \frac{\Delta V}{V} = e_{xx} + e_{yy} + e_{zz} =
         - (2 S_{11} + 2 S_{12} + 4 S_{13} + S_{33}) P  \; ,
\end{equation}
and using the relations between stiffness and compliance 
constants,\cite{ma18,ra20b} we find
\begin{equation}
 B = - V \frac{\partial P}{\partial V} =
     \frac { (C_{11} + C_{12}) \, C_{33} - 2 C_{13}^2 } 
           { C_{11} + C_{12} + 2 C_{33} - 4 C_{13} }    \; ,
\label{bulkm2}
\end{equation}
as in Refs.~\onlinecite{ja87,bo97}.

\begin{figure}
\vspace{-0.6cm}
\includegraphics[width=7cm]{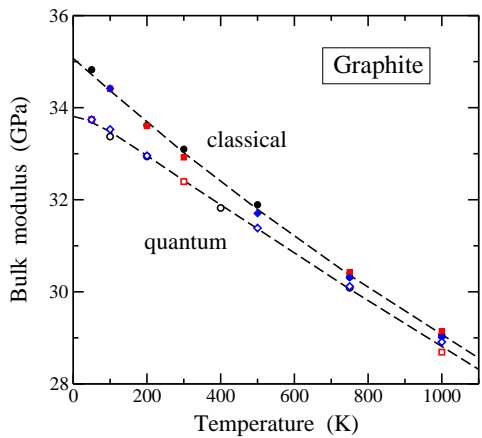}
\vspace{-0.5cm}
\caption{Temperature dependence of the bulk modulus of graphite,
obtained from PIMD (open symbols) and classical MD simulations
(solid symbols). The data shown in both cases were obtained by
following three different procedures:
(1) numerical derivative $d P / d V$ from various hydrostatic
pressures $P$ (squares), (2) fluctuation formula in Eq.~(\ref{bulkm})
(circles), and (3) from the elastic constants using Eq.~(\ref{bulkm2})
(diamonds).
Error bars are on the order of the symbol size.
Dashed lines are guides to the eye.
}
\label{f10}
\end{figure}

The temperature dependence of the bulk modulus $B$ of graphite
derived from our simulations is shown in Fig.~10. Solid and
open symbols represent results of classical and quantum
simulations, respectively.
In each case, symbols represent data obtained from
(1) numerical derivative $d P / d V$ (squares),
(2) fluctuation formula (circles), and (3) elastic constants
(diamonds).
The values obtained for the bulk modulus from the three methods
agree well in the whole temperature range considered here,
for both quantum and classical data.
In the limit $T \to 0$, we find $B$ = 35.1 and 33.8 GPa for
the classical and quantum models, respectively. This means
a reduction of the bulk modulus by a 4\% due to zero-point
motion of the C atoms.

In Table~III we give values of the bulk modulus $B$ of graphite
calculated using Eq.~(\ref{bulkm2}) from the stiffness constants 
derived from our 
classical and quantum simulations at $T$ = 300 and 750~K, 
as well as for the zero-temperature limit. 
Values of $B$ obtained from the elastic constants yielded
by experimental and theoretical methods are given in Tables~IV
and V, respectively.
From an analysis of the equation of state of graphite at
room temperature, Hanfland {\em et al.}\cite{ha89} found
at ambient pressure $B$ = 33.8(30) GPa, a little higher than 
our quantum result at 300~K ($B$ = 32.3 GPa).
Zhao and Spain\cite{zh89} obtained a somewhat larger value,
$B$ = 35.8(16), from x-ray diffraction experiments.
Tohei {\em et al.}\cite{to06} found $B$ = 28.7 GPa at 300~K from 
LDA-DFT calculations combined with a QHA for the lattice modes.

One can also define an ``isotropic bulk modulus'' $B_{\rm iso}$
for isotropic changes of the volume,\cite{ja87} which means 
$e_{xx} = e_{yy} = e_{zz} = e$ and $\Delta V / V = 3 e$.  
One thus has a hydrostatic pressure 
\begin{equation}
  P = - \frac13 (\tau_{xx} + \tau_{yy} + \tau_{zz})  \; ,
\end{equation}
which gives
\begin{equation}
 B_{\rm iso} =   - V \frac{\partial P}{\partial V} =
  \frac19 \, \left( 2 C_{11} + 2 C_{12} + 4 C_{13} + C_{33} \right) \; .
\label{bulkm_iso}
\end{equation}
For the classical limit at $T = 0$ we have $B_{\rm iso}$ = 276.6~GPa,
and from our PIMD simulations we find for $T \to 0$:
$B_{\rm iso}$ = 263.5~GPa.
At $T$ = 300 K, the classical and quantum simulations yield
259.6 and 254.6~GPa, respectively.

\section{Summary}

PIMD simulations allow us to quantify nuclear quantum effects in
structural and elastic properties in condensed matter. 
For graphite, in particular, we have seen that such quantum 
effects are appreciable for $T$ in the order of 500~K, and even
higher for some variables.
At low temperature, the quantum zero-point expansion of the graphite 
volume is nonnegligible, and amounts to 1.3\% of the classical value.
In spite of the large anisotropy of graphite, we have found that
the expansion due to zero-point motion is nearly isotropic, i.e., 
the relative increases in in-plane and out-of-plane directions
are approximately the same.

The thermal contraction of the in-plane area ($\alpha_x < 0$) 
observed in x-ray diffraction experiments at low temperature
is reproduced by our quantum simulations, contrary to classical
MD, where a positive in-plane thermal expansion is found in 
the whole temperature range studied here. 
Given that a negative $\alpha_x$ 
in layered materials is caused by out-of-plane atomic motion,
the in-plane contraction of graphite is essentially due to 
quantum motion of the C atoms in the $z$-direction.
Also, the characteristic trend of $\alpha_x$ (negative at low $T$
and positive at high $T$) is a clear signature of anharmonicity
in the vibrational modes, indicating a coupling between in-plane 
and out-of-plane modes.

Quantization of lattice vibrations gives rise to changes in the 
elastic properties of graphite with respect to a classical model.
At low temperature, the most significant relative changes in 
the elastic stiffness constants correspond to $C_{12}$ and 
$C_{44}$, where quantum corrections cause a reduction of 
19\% and 28\%, respectively.
The bulk modulus and Poisson's ratio decrease by a 4\% and
19\% at low $T$ because of zero-point motion of the carbon atoms.
In general, our results indicate that graphite is ``softer''
than predicted by classical simulations.

In connection with $C_{13}$, it is still
open the question why several {\em ab-initio} calculations have
yielded negative values, which has not been observed in
experimental studies. We found here positive values for this
elastic constant, but we did not observe any quantum effect
on it. All this could be due to a lack of precision in the
description of interlayer van-der-Waals-like interactions. 

We finally note the consistency of the simulation results with 
the third law of thermodynamics.
This means, in particular, that for $T \to 0$, thermal expansion 
coefficients should vanish. Moreover, the temperature 
derivatives of the elastic stiffness constants and bulk modulus
should also vanish.

\begin{acknowledgments}
This work was supported by Ministerio de Ciencia e Innovaci\'on
(Spain) through Grant PGC2018-096955-B-C44.
\end{acknowledgments}


\end{document}